# Measurables of $CP$ Violation in $B_d \to D^0_{CP} K_S$ at a $B$-meson Factory

Aik Hui CHAN

*Nanyang Polytechnic, French-Singapore Institute,*

*12 Science Centre Road, Singapore 2260*

Zhi-zhong XING [1]

*Sektion Physik, Theoretische Physik, Universität München,*

*Theresienstrasse 37, D-80333 Munich, Germany*

**Abstract**

In the context of the standard electroweak model, we emphasize that $B_d \to D^0_{CP} K_S$ ($D^0_{CP}$ denotes a $CP$ eigenstate of $D^0$ or $\bar{D}^0$) can compete with $B_d \to \pi^+\pi^-$ in studying $CP$ violation and probing the Cabibbo-Kobayashi-Maskawa unitarity triangle. We discuss the measurables of direct and indirect $CP$ asymmetries in $B^0_d$ vs $\bar{B}^0_d \to D^0_{CP} K_S$ under the circumstance of an asymmetric $B$-meson factory running on the $\Upsilon(4S)$ resonance, and show that both the weak and strong phases are experimentally determinable even in the presence of unknown final-state interactions.

Today the most promising way to test the Cabibbo-Kobayashi-Maskawa (CKM) mechanism of quark mixing and $CP$ violation is to measure $CP$ asymmetries in neutral $B$-meson decays to $CP$ eigenstates [1]. Such measurements are becoming feasible with the development of the $B$ factory programs [2]. In addition to $B_d \to J/\psi K_S$, $B_d \to \pi^+\pi^-$ is regarded as another interesting and feasible process for the study of $CP$ violation and in particular for extraction of the CKM phase(s). It is known that $B_d \to \pi^+\pi^-$ has a branching ratio of order $10^{-5}$ and its $CP$ asymmetry suffers from the uncertainties induced by penguin diagrams [3]. In contrast, we find that the decay mode $B_d \to D^0_{CP} K_S$ ($D^0_{CP}$ denotes a $CP$ eigenmode of $D^0$ or $\bar{D}^0$) occurs only through the tree-level quark diagrams (see Fig. 1) and its branching ratio is also of order $10^{-5}$ [4]. Hence it is expected that $B_d \to D^0_{CP} K_S$ can practically compete with $B_d \to \pi^+\pi^-$ in probing $CP$ violation and testing the CKM unitarity triangle.

Although some attention [5] has been paid to $B_d^0 \to \overset{(-)}{D}{}^{(*)0} K_S$ and their $CP$-conjugate processes for extraction of the third angle of the CKM unitarity triangle ($\gamma$), a study of the relevant observables of $CP$ violation at the accessible experimental scenarios has been lacking. The purpose of this short note is to instructively analyze the measurables of direct and indirect $CP$ asymmetries in $B_d^0$ vs $\bar{B}_d^0 \to D^0_{CP} K_S$ under the circumstance of an asymmetric $B$ factory running on the $\Upsilon(4S)$ resonance. We show that both the weak and strong phases in these decay modes are in principle determinable provided that the $CP$ asymmetries are measured. The signal of direct $CP$ violation can be distinguished from that of indirect $CP$ violation in the time-dependent measurements.

$CP$ violation in the $K^0 - \bar{K}^0$ and $D^0 - \bar{D}^0$ systems is negligible in comparison with that in the $B_d^0 - \bar{B}_d^0$ system. Hence one can use $K_S = (K^0 + \bar{K}^0)/\sqrt{2}$ to denote the $CP$-even state of neutral $K$ mesons, and $D^0_{(\pm)} = (D^0 \pm \bar{D}^0)/\sqrt{2}$ to denote the $CP$-even (+) and $CP$-odd (−) states of neutral $D$ mesons [2]. In terms of the inner angles of the CKM unitarity triangle [6], the decay amplitudes of $B_d^0$ vs $\bar{B}_d^0 \to D^0_{(\pm)} K_S$ are parametrized as follows:

$$\begin{aligned}
\langle D^0_{(\pm)} K_S | \mathcal{H} | B_d^0 \rangle &= \tfrac{1}{2}\left[\langle D^0 K^0 | \mathcal{H} | B_d^0 \rangle \pm \langle \bar{D}^0 K^0 | \mathcal{H} | B_d^0 \rangle\right] = \tfrac{1}{2}\left(e^{i\gamma} A_x \pm A_y\right), \\
\langle D^0_{(\pm)} K_S | \mathcal{H} | \bar{B}_d^0 \rangle &= \tfrac{1}{2}\left[\langle D^0 \bar{K}^0 | \mathcal{H} | \bar{B}_d^0 \rangle \pm \langle \bar{D}^0 \bar{K}^0 | \mathcal{H} | \bar{B}_d^0 \rangle\right] = \tfrac{1}{2}\left(A_y \pm e^{-i\gamma} A_x\right),
\end{aligned} \quad (1)$$

where the complex hadronic amplitudes $A_x$ and $A_y$ contain the real CKM factors as well as the strong phases due to final-state interactions. The ratio of the above transition amplitudes

---
[2] We have used the phase convention $CP|P^0\rangle = |\bar{P}^0\rangle$ and $CP|\bar{P}^0\rangle = |P^0\rangle$, where $P = K$, $D$ or $B_d$.

is given by

$$\xi_\pm = \frac{\langle D^0_{(\pm)} K_S | \mathcal{H} | \bar{B}^0_d \rangle}{\langle D^0_{(\pm)} K_S | \mathcal{H} | B^0_d \rangle} = \frac{\chi e^{i\delta} \pm e^{-i\gamma}}{e^{i\gamma} \pm \chi e^{i\delta}} , \qquad (2)$$

where $\chi = |A_y/A_x|$, and $\delta$ stands for the phase shift between $A_x$ and $A_y$. Note that $\delta$ is in general nonvanishing, since the final states $D^0 K^0$ (or $\bar{D}^0 \bar{K}^0$) and $\bar{D}^0 K^0$ (or $D^0 \bar{K}^0$) are different in the isospin configuration. The presence of nonzero $\delta$ signifies direct $CP$ violation. Regardless of final-state interactions, one can estimate the size of $\chi$ with the help of the factorization approximation. Because the four decay modes $\overset{(-)}{B}{}^0_d \to \overset{(-)}{D}{}^0 \overset{(-)}{K}{}^0$ have the same decay constants and formfactors, we obtain $\chi \approx 1/\sqrt{\rho^2 + \eta^2} \sim 3$ (here $\rho$ and $\eta$ are the Wolfenstein parameters [7]).

At an asymmetric $B$-meson factory running on the $\Upsilon(4S)$ resonance, the $B$'s are produced in a two-body ($B^+_u B^-_u$ or $B^0_d \bar{B}^0_d$) state with *odd* charge-conjugation parity. Since the two neutral $B$ mesons mix coherently untill one of them decays, one can use the semileptonic decay of one meson to tag the flavour of the other meson decaying to a hadronic $CP$ eigenstate. The (proper) time distribution of the decay rates for $B^0_d$ vs $\bar{B}^0_d \to D^0_{(\pm)} K_S$ can be given by [8]

$$\begin{aligned} R(l^-, D^0_{(\pm)} K_S; t) &\propto e^{-\Gamma|t|} \left[ 1 + |\xi_\pm|^2 + (1 - |\xi_\pm|^2) \cdot \cos(x\Gamma t) - 2\mathrm{Im}\left(e^{-2i\beta}\xi_\pm\right) \cdot \sin(x\Gamma t) \right] , \\ R(l^+, D^0_{(\pm)} K_S; t) &\propto e^{-\Gamma|t|} \left[ 1 + |\xi_\pm|^2 - (1 - |\xi_\pm|^2) \cdot \cos(x\Gamma t) + 2\mathrm{Im}\left(e^{-2i\beta}\xi_\pm\right) \cdot \sin(x\Gamma t) \right] , \end{aligned} \qquad (3)$$

where $t$ is the time difference between the semileptonic and nonleptonic decays [3], $x \equiv \Delta m / \Gamma \approx 0.71$ denotes a mixing parameter of the $B^0_d - \bar{B}^0_d$ system, and $\beta$ is another angle of the CKM unitarity triangle [6]. From Eq. (3) the time-dependent $CP$ asymmetries are obtained as

$$\mathcal{A}_\pm(t) = \frac{R(l^-, D^0_{(\pm)} K_S; t) - R(l^+, D^0_{(\pm)} K_S; t)}{R(l^-, D^0_{(\pm)} K_S; t) + R(l^+, D^0_{(\pm)} K_S; t)} = \mathcal{U}_\pm \cos(x\Gamma t) - \mathcal{V}_\pm \sin(x\Gamma t) , \qquad (4a)$$

where the subscripts "$\pm$" correspond to the final states $D^0_{(\pm)} K_S$, and

$$\mathcal{U}_\pm = \frac{1 - |\xi_\pm|^2}{1 + |\xi_\pm|^2} , \qquad \mathcal{V}_\pm = \frac{2\mathrm{Im}\left(e^{-2i\beta}\xi_\pm\right)}{1 + |\xi_\pm|^2} . \qquad (4b)$$

Clearly the terms $\mathcal{U}_\pm$ and $\mathcal{V}_\pm$ in $\mathcal{A}_\pm(t)$, which are distinguishable in the time-dependent measurements, represent direct and indirect $CP$ violation respectively.

---

[3]Note that the time sum of the semileptonic and nonleptonic decays has been integrated out, since it is not measured at a $B$ factory [2,8].

In terms of the weak- and strong-interaction parameters, we re-express the measurables of direct and indirect $CP$ asymmetries as follows:

$$\begin{aligned}\mathcal{U}_{\pm} &= \frac{\pm 2\chi \sin\gamma \sin\delta}{1+\chi^2 \pm 2\chi \cos\gamma \cos\delta} ,\\ \mathcal{V}_{\pm} &= \frac{\mp \sin[2(\beta+\gamma)] - 2\chi \cos\delta \sin(2\beta+\gamma) \mp \chi^2 \sin(2\beta)}{1+\chi^2 \pm 2\chi \cos\gamma \cos\delta} .\end{aligned} \quad (5)$$

One can observe that in the above four measurables ($\mathcal{U}_{\pm}$ and $\mathcal{V}_{\pm}$) there only exist two weak-interaction parameters ($\beta$ and $\gamma$) and two strong-interaction parameters ($\chi$ and $\delta$). Thus it is possible to determine all theses four parameters provided that the measurements of $B^0_d \to D^0_{(\pm)} K_S$ and $\bar{B}^0_d \to D^0_{(\pm)} K_S$ are realized at $B$ factories. This also implies that one can extract the CKM phase(s) unambiguously, even in the presence of final-state interactions. If $\delta$ happens to be very small (or vanishing), $\mathcal{U}_{\pm}$ may be suppressed to the level that is not detectable. In this case, $\chi \approx 1/\sqrt{\rho^2+\eta^2}$ is a safe result and $\mathcal{V}_{\pm}$ becomes

$$\mathcal{V}_{\pm}(\delta \approx 0) \approx \frac{\mp \sin[2(\beta+\gamma)] - 2\chi \sin(2\beta+\gamma) \mp \chi^2 \sin(2\beta)}{1+\chi^2 \pm 2\chi \cos\gamma} . \quad (6)$$

Since $\chi$ is well determinable from measuring $|V_{ub}|/|V_{cb}|$, the weak phases $\beta$ and $\gamma$ can still be extracted from $\mathcal{V}_{\pm}(\delta \approx 0)$. At present the experimental constraints to these two angles are $5^0 \leq \beta \leq 45^0$ and $20^0 \leq \gamma \leq 165^0$ [9]. For illustration, we estimate the sizes of $\mathcal{U}_{\pm}$ and $\mathcal{V}_{\pm}$ as functions of the strong phase shift $\delta$ in Fig. 2, where $\chi \approx 3$, $\beta \approx 18^0$ and $\gamma \approx 72^0$ are typically taken. Clearly the effect of $\delta$ on the $CP$-violating measurables is significant.

In a similar way one can discuss $CP$ violation in $B^0_d$ vs $\bar{B}^0_d \to D^0_{CP} K_L$, where $K_L = (K^0 - \bar{K}^0)/\sqrt{2}$. It is easy to show the following relations:

$$\mathcal{U}_{\pm}(D^0_{CP} K_L) = \mathcal{U}_{\pm}(D^0_{CP} K_S), \qquad \mathcal{V}_{\pm}(D^0_{CP} K_L) = -\mathcal{V}_{\pm}(D^0_{CP} K_S) . \quad (7)$$

In the context of the CKM mechanism, one or two of the above $CP$ asymmetries may reach the level of 50% (see Fig. 2). To establish such a signal of $CP$ violation up to 3 standard deviations at an asymmetric $B$ factory, about $10^{8-9}$ $B^0_d \bar{B}^0_d$ pairs (including the cost for flavour tagging) are needed. Of course this number of events is accessible in the near future.

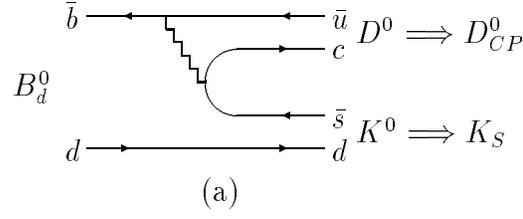

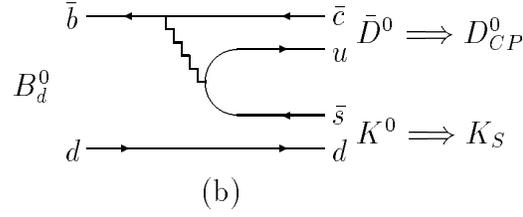

Figure 1: Quark diagrams for a $B_d^0$ meson decaying to the $CP$ eigenstate $D_{CP}^0 K_S$.

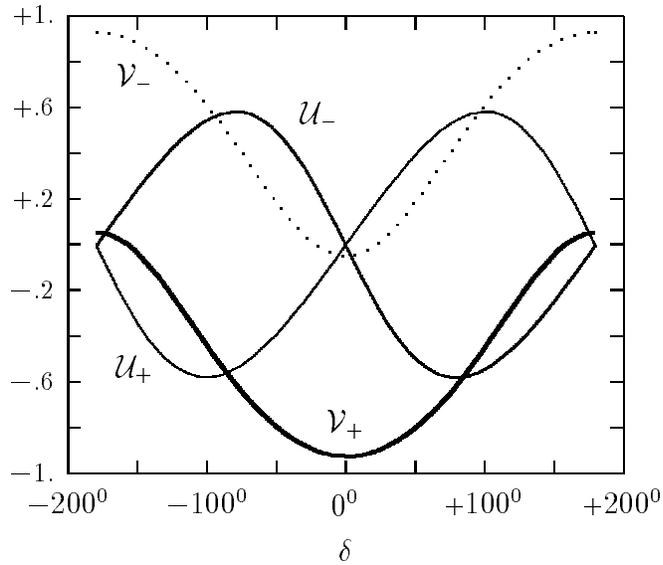

Figure 2: Measurables of direct and indirect $CP$ violation ($\mathcal{U}_\pm$ and $\mathcal{V}_\pm$) as functions of the strong phase shift ($\delta$) in $B_d^0$ vs $\bar{B}_d^0 \to D_{(\pm)}^0 K_S$ at asymmetric $B$-meson factories.

6